# Exclusive measurement of two-pion production in the $dd \rightarrow {}^4\text{He}\,\pi\pi$ reaction

S. Keleta [a], Chr. Bargholtz [b], M. Bashkanov [c], M. Berłowski [d], D. Bogoslawsky [e], H. Calén [a], H. Clement [c], L. Demirörs [f], C. Ekström [g], K. Fransson [a], L. Gerén [b], L. Gustafsson [a], B. Höistad [a,*], G. Ivanov [e], M. Jacewicz [a], E. Jiganov [e], T. Johansson [a], O. Khakimova [c], F. Kren [c], S. Kullander [a], A. Kupść [a], A. Kuzmin [h], K. Lindberg [b], P. Marciniewski [a], B. Morosov [e], W. Oelert [i], C. Pauly [f], H. Petrén [a], Y. Petukhov [e], A. Povtorejko [e], A. Pricking [c], R.J.M.Y. Ruber [a], K. Schönning [a], W. Scobel [f], R. Shafigullin [j], B. Shwartz [h], T. Skorodko [c], V. Sopov [k], J. Stepaniak [d], P.-E. Tegnér [b], P. Thörngren Engblom [b], V. Tikhomirov [e], A. Turowiecki [l], G.J. Wagner [c], C. Wilkin [m], M. Wolke [i], J. Zabierowski [n], I. Zartova [b], and J. Złomańczuk [a]

[a] Department of Physics and Astronomy, Uppsala University, Box 535, S-75121 Uppsala, Sweden
[b] Department of Physics, Stockholm University, S-10191 Stockholm, Sweden
[c] Physikalisches Institut der Universität Tübingen, D72076 Tübingen, Germany
[d] Soltan Institute of Nuclear Studies, PL-00681 Warsaw, Poland
[e] Joint Institute for Nuclear Research, Dubna, 101000 Moscow, Russia
[f] Institut für Experimentalphysik der Universität Hamburg, Hamburg, D-22761 Hamburg, Germany
[g] The Svedberg Laboratory, S-75121 Uppsala, Sweden
[h] Budker Institute of Nuclear Physics, Novosibirsk 630 090, Russia
[i] Institut für Kernphysik, Forschungszentrum Jülich GmbH, 52425 Jülich, Germany
[j] Moscow Engineering Physics Institute, Moscow, Russia
[k] Institute of Theoretical and Experimental Physics, Moscow, Russia
[l] Institute of Experimental Physics of Warsaw University, PL-00681 Warsaw, Poland
[m] Physics and Astronomy Department, UCL, London WC1E 6BT, United Kingdom
[n] Soltan Institute of Nuclear Studies, PL-90137 Lodz, Poland

**Abstract**

The results from the first kinematically complete measurement of the $dd \rightarrow {}^4\text{He}\,\pi\pi$ reaction are reported. The aim was to investigate a long standing puzzle regarding the origin of the peculiar $\pi\pi$−invariant mass distributions appearing in double pion production in light ion collisions, the so-called ABC effect. The measurements were performed at the incident deuteron energies of 712 MeV and 1029 MeV, with the WASA detector assembly at CELSIUS in Uppsala, Sweden. We report the observation of a characteristic enhancement at low $\pi\pi$-invariant mass at 712 MeV, the lowest energy yet. At the higher energy, in addition to confirming previous experimental observations, our results reveal a strong angular dependence of the pions in the overall centre of mass system. The results are qualitatively reproduced by a theoretical model, according to which the ABC effect is described as resulting from a kinematical enhancement in the production of the pion pairs from two parallel and independent $NN \rightarrow d\,\pi$ sub-processes.

---

*Corresponding author.

E-mail address: bo.hoistad@fysast.uu.se (B. Höistad).

# 1. Introduction

In the 1960s, experiments with proton beams colliding with deuterium targets produced unexpected results in the momentum spectra of the observed $^3$He particles [1]. This anomaly, later called the ABC effect after the initials of its discoverers, refers to the unexpected enhancement in the momentum spectrum of the recoiling nuclei corresponding to a missing mass a little above the two-pion threshold. Neither the suggestions given at the time, nor those that followed later, could explain all manifestations of the effect satisfactorily; even what constituted the ABC effect was sometimes cast into doubt one way or another. The basic issue to resolve is whether the ABC effect is a reaction mechanism phenomenon, or if it has its origin in some final state interaction or in any other dynamic phenomenon.

The main features of the ABC effect comprise:
- it manifests itself as an enhancement in the invariant mass distribution of two pions, typically at ~ 310 MeV/c$^2$, with a width of about 50 MeV/c$^2$,
- the accompanying outgoing nucleons should be bound together or have low excitation energy,
- it does not occur when the two pions are created in isospin one, and hence it is an even angular momentum state.

The isospin identification is made clear by the comparison of the missing-mass spectra from the $pd \to {}^3\text{He}\,X$ and $pd \to {}^3\text{H}\,X$ reactions where, in the latter case, the absence of an enhancement is explained by the $2\pi$-system $X$ having isospin $I_x = 1$ [1]. The study of the ABC effect in the $pd \to {}^3\text{He}\,X$ channel, often accomplished using a deuteron beam, was pursued by several groups [2-4]. At low excess energies the ABC enhancement was little seen and the effect seemed to be largest at energies where the maximum missing mass was around 500-600 MeV/c$^2$. By far the most detailed investigation in this region was reported in Ref. [5], where kinematically complete measurements of the $pd \to {}^3\text{He}\,\pi^+\pi^-$ and $pd \to {}^3\text{He}\,\pi^0\pi^0$ reactions were undertaken at $T_p$= 893 MeV, which corresponds to the $\Delta\Delta$-excitation region, where in previous inclusive measurements the maximum of the ABC effect has been observed.

The ABC effect has also been studied with mixed results in the $np \to d\,X$ reaction, though here the experimental situation is not as clear-cut because of the use of a broad-band neutron beam or a neutron target moving with Fermi momentum in a deuterium target [6-10]. To overcome this problem, exclusive measurements of $pd \to p_{sp}d\pi^0\pi^0$ have been carried out [11]. By measuring all the final particles apart from the "spectator" proton ($p_{sp}$), the total centre-of-mass energy (cm) could be established on an event-by-event basis and the effects of Fermi smearing eliminated. Clear ABC peaks were then observed at beam energies of 1.03 and 1.35 GeV. The ABC enhancement is also evident in the $pp \to ppX$ reaction, provided that the final diproton has low excitation energy [12]. Although the two pions in the peak are here in an $s$-wave, it is possible that there could be an $I = 2$ component in this reaction.

The cleanest channel in which to study the ABC would seem to be through double-pion production in the $dd \to {}^4\text{He}\,\pi\pi$ reaction, since here the pion-pair is produced in a pure isospin-zero state. This was investigated at the Saturne accelerator, where inclusive measurements were conducted for fixed $^4$He production angles at several deuteron beam energies, ranging from 0.8 to 2.4 GeV [13,14]. The $^4$He momentum spectra show a spectacular ABC structure with peaks at both small and large $^4$He momenta, corresponding to the forward and backward production of the low mass enhancement in the cm system. In addition there was always a broad bump in the middle of the spectrum corresponding to an enhancement at the maximum allowed missing mass.

The situation is very different at low energy, where measurements at 649 MeV [15] and very close to threshold at 570 MeV [16,17] showed low cross sections and no ABC enhancement, with distributions that were broadly consistent with phase space predictions.

Several ideas have been put forward to explain the ABC effect. The idea that this enhancement could be due to a new particle or resonance was ruled out early since its position and width changed with the kinematic conditions. Neither could it be due to a large s-wave $\pi\pi$-scattering length, as suggested at the time by the original authors [1]. The $\pi\pi$-scattering length required was larger by an order of magnitude than expected on theoretical grounds [18] or determined experimentally [19].

Some of the theoretical models put forward to explain the ABC phenomenon include the $\Delta\Delta$-excitation mechanism of Risser and Shuster [20], the double-nucleon-exchange model of Anjos et al. [21], the chiral bag model of Kälbermann and Eisenberg [22], and the more general approach of Alvarez-Ruso et al. to all $NN \to NN\pi\pi$ reaction channels [23,24]. The double-$\Delta$ approach was extended through the inclusion of ρ-meson exchange [25] and through the introduction of a $\Delta\Delta$ final state interaction [26].

The most successful dynamical model for the ABC phenomenon is that proposed by Gårdestig, Fäldt, and Wilkin (GFW) to describe the inclusive $dd \to {}^4\text{He}\,X$ reaction [27]. This is a model of the double-$\Delta$ type where the $dd \to {}^4\text{He}\,\pi\pi$ is treated as being driven by two independent $NN \to d\pi$ reactions occurring in parallel. Not only does this model reproduce quantitatively the missing-mass spectra obtained between 0.8 and 2.4 GeV [13], it also predicts quite well the measured deuteron analyzing powers [14]. On the other hand, both this and other dynamical models [28] fail completely to describe the low energy data [15-17].

Since all previous $dd$ experiments were inclusive, with only the $^4$He being detected, it was suggested that all outgoing particles ($^4$He and pions) should be measured with the WASA detector at CELSIUS [29,30]. This would provide for the first time angular distributions for all final particles, as well as two-particle invariant mass distributions for all combinations. We present kinematically complete measurements of both $dd \to {}^4\text{He}\,\pi^+\pi^-$ and $dd \to {}^4\text{He}\,\pi^0\pi^0$, in two crucial energy regions, one where there has been no previous measurement, and another where exclusive measurements were lacking, but where the ABC effect is known to be prominent.

The WASA experimental facility, as installed at CELSIUS, is described in section 2, with the data analysis being discussed in some detail in the following section. The results obtained for $dd \to {}^4\text{He}\,\pi^+\pi^-$ and $dd \to {}^4\text{He}\,\pi^0\pi^0$ at 1029 MeV are presented in section 4. Our conclusions are drawn in section 5.

## 2. The experimental set-up

The experiment was carried out at the CELSIUS storage ring at the The Svedberg Laboratory in Uppsala, Sweden, using WASA, the Wide Angle Shower Apparatus. This 4π detector was designed for the study of the production and decay of light mesons ($\pi^o$, η, ω...) produced in collisions of light nuclei. At CELSIUS energies, these light mesons, except for the pion, can only be produced near threshold. The outgoing nucleons are therefore confined to within a narrow cone around the forward direction, while the meson decay products are spread out over all directions in the laboratory system. The WASA geometry and its internal target design are well optimized for this particle emission configuration. The detector was located on one of the straight sections of the CELSIUS ring until the summer of 2005. It was then relocated to COSY–Jülich, where it has been operating since the autumn of 2006. A cross section of the central and forward parts of WASA is shown in. Fig. 1.

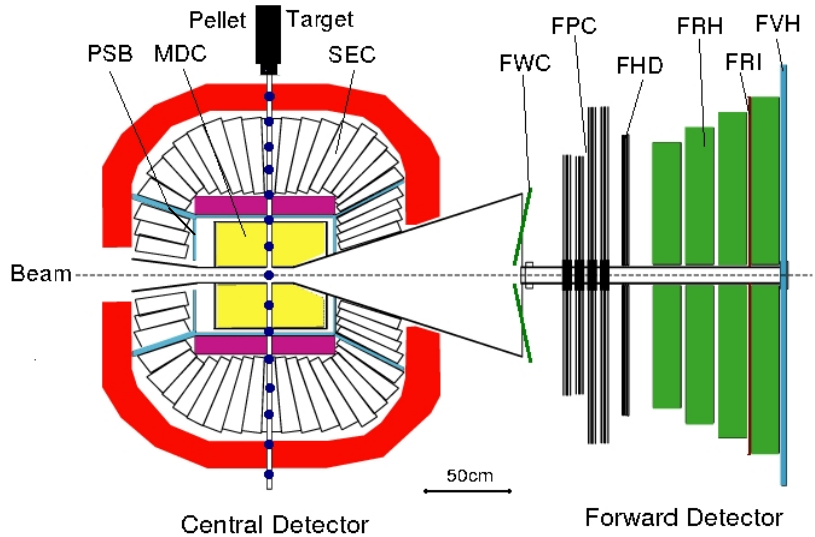

Fig. 1. Cross section of the WASA detector.

The WASA facility has three major components: the Forward Detector together with a Tagging Spectrometer (not shown in the figure) for measuring the outgoing nuclei, the Central Detector for measuring meson decay products, and the internal pellet target system. A comprehensive account of the WASA detector can be found in Ref. [31] and we will here be brief.

*2.1 The forward detector*

The forward detector is mainly designed for the measurement of the energy and direction of charged reaction products such as protons, deuterons and He from collisions of light nuclei. The range of polar angles is $3°$ to $17°$ around the beam direction. It has several detector planes with different and complementary functions. The first is the Forward Window Counter (FWC), whose main purpose is to reduce background from particles scattered on the beam pipe and from other secondary interactions on the support structures in the forward detector. The second consists of the Forward Proportional Chambers (FPC) used for particle track reconstruction. The third is made up of three layers of plastic scintillators (FHD) used to determine the scattering and azimuthal angles of a particle track. It also forms part of the trigger for on-line track selection. Then follow the planes of the Forward Range Hodoscope (FRH) used to measure energy deposited by charged particles. The first two layers were also included in the trigger for on-line event selection. The multiple layers, together with the third layer of FHD, can also be used for $\Delta E$-$E$ particle identification. There is also the FRI detector, a mesh of plastic scintillators, for neutron track reconstruction in the forward detector. The final plane of the forward detector is the Forward Veto Hodoscope (FVH), used for on-line event selection as part of the trigger system and for off-line proper energy reconstruction of punch-through particles.

*2.2 The central detector*

The central detector was designed to detect photons as well as charged particles (electrons, pions) from the decay of light mesons. Its main components are the mini-drift chamber (MDC), the plastic scintillator barrel (PSB) and the scintillating electromagnetic calorimeter (SEC). The SEC, which is a major part of the central detector, consists of 1012 CsI (Na) crystals forming 24 rings around the beam pipe. A super conducting solenoid is positioned between the PSB and the SEC.

*2.3 The pellet target*

The WASA pellet system generates frozen hydrogen or deuterium pellets to be used as targets with minimal inactive material around them in the interaction region. The pellets are produced from a stream of liquid target material (hydrogen or deuterium) that is broken into droplets by a vibrating nozzle. The approximately 35 μm diameter droplets are then frozen by evaporation, collimated and injected into the scattering chamber. After passing through the target region, the pellets are collected in a pellet dump.

**3. Data analysis**

The aim of measuring the $dd \rightarrow {}^4\text{He}\pi\pi$ reaction was to provide data for an exclusive investigation of the ABC-effect in the energy region between previous near-threshold measurements that showed little deviation from phase space [15-17], up to the energies characteristic of the ABC effect, where only inclusive data existed [13,14]. The measurements were carried out at two beam energies, specifically $T_d$= 712 MeV and $T_d$ = 1029 MeV.

A dedicated He trigger, based on coincident hits in two layers of the forward detector (FHD3+FRH1), was developed in order to select events that included He candidates in the forward detector. In order to have an unbiased inclusive data set, no conditions were placed on the central detector. This minimal requirement allowed the recording of all reaction channels involving neutral and charged pions in the central detector. As a result, a total of 76 700 $dd \rightarrow {}^4\text{He}X$ events were collected by the forward detector at 1029 MeV, including 8 100 $dd \rightarrow {}^4\text{He}\pi^o\pi^o$ and 8 600 $dd \rightarrow {}^4\text{He}\pi^+\pi^-$ identified events, where the central detector was used for the detection of the pions. At 712 MeV, approximately 3 600 $dd \rightarrow {}^4\text{He}\pi^o\pi^o$ events were obtained.

*3.1 Overall calibration*

The overall calibration procedure in this context refers to the process of converting the ADC read-out to deposited energy, which in our case involved: gain calibration, correction for time-drift of the gain, and correction for non-uniformity of the light collection. The gain calibration has linear and non-linear components which are dealt with simultaneously. These corrections are usually done by comparing real data with the results of Monte Carlo simulations.

Gain drift correction refers to the variation of the gain in the detectors over time during a given run period. This could have adverse effects on particle identification and energy reconstruction. Gain drift over time can be checked by using fast particles that are expected to deposit the same amount of energy per detector or using a light-pulser signal. In our case, even though there was a noticeable drift of the gain, the deviation was too small to have an effect on particle identification. Nevertheless, the correction factors obtained from light-pulser signals were implemented.

The calibration of the forward detector was done by comparing the standard Δ*E*-Δ*E* plots between each pair of the forward detector planes with full GEANT Monte Carlo simulations of the detector. The elements involved in the calibration are the third plane of the FHD (FHD3) and all four planes of the FRH. These are all pie-shaped plastic scintillators. As a result, a clean separation of ³He and ⁴He was achieved in the forward detector planes for all FRH1 and FRH2 detectors. For FHD3, a reasonable separation could be obtained for only about half of them, due to aging effects in many of the detector elements.

The light output from a scintillator depends, not only on the amount of energy deposited, but also on the position of impact relative to the photomultiplier tube, i.e., on the production angle of the particle in question. Corrections for such dependence can be derived from fast particles that deposit nearly constant energy for all production angles. In general, when the non-uniformity corrections thus obtained were not good enough, new values were derived using the so-called band-projection method outlined in Ref. [32]. An example of the improvement that can be achieved by this method is shown in Fig. 2.

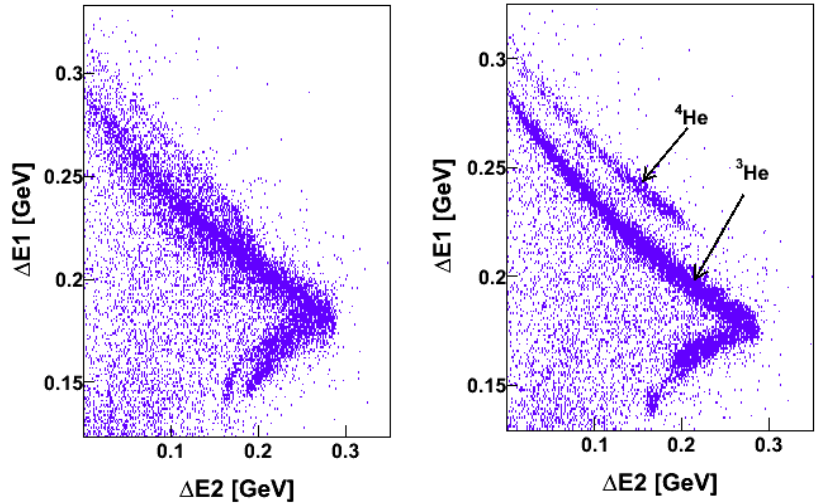

Fig. 2. An example of a $\Delta E$-$\Delta E$ plot at $T_d$=1029 MeV for one pair of FRH (FRH1 versus FRH2), where the non-uniformity was severe; left panel is before and the right one after correction. An improved separation of $^3$He and $^4$He is then obtained as seen from the figures.

Since the energy resolution of the FHD3 detectors was not satisfactory, the separation of $^3$He and $^4$He bands could be achieved for only about half of the 48 elements and only the good elements were used in the final analysis. Furthermore, individual gates corresponding to these detectors were applied on the $\Delta E$-$\Delta E$ plots in order to avoid particle misidentification. The FRH detectors posed no such problems.

*3.2 Event reconstruction*

The event reconstruction process involved particle track reconstruction, particle identification and kinetic energy reconstruction.

*3.2.1 Track reconstruction*
The criteria applied for a good track candidate were in the forward detector:
- no missing plane in a track until it stops,
- the production angle has to be between 3° and 17°,
- for the $dd \to {}^4\text{He}\pi\pi$ reaction, the track must reach at least the first plane of FRH for the 712 MeV data, and the second plane for the 1029 MeV data.

In the central detector the criteria were:
- presence of a hit cluster in the calorimeter,
- presence (absence) of a plastic barrel signal for a charged (neutral) track,
- the production angle has to be between 20° and 145°.

*3.2.2 Particle identification*
With regards to particle identification, an event of interest must contain a $^4$He particle and two pions in the final state. These are identified as follows. In the forward detector, particle identification is done using the $\Delta E$-$\Delta E$ method between the forward detector planes. In the central detector, the neutral pions are identified from their decay to two photons, each of which is identified by a central detector track with no plastic barrel signal. Charged pions are identified from their trajectory in the mini-drift chamber (MDC), in the presence of a magnetic field.

*3.2.3 Deposited-to-kinetic-energy translation*

At a beam energy of 1029 MeV, $^3$He and $^4$He particles from the $dd \to \text{He} X$ reaction stop at different depths in the forward detector planes (FRH planes), the fastest $^4$He reaching the second layer of FRH. Because of losses in detector support structures, the deposited energy of the particles is less than the incident kinetic energy. One therefore has to determine the kinetic energy from the deposited energies, taking into account these and other losses.

The kinetic energy of He is derived from the deposited energy in the layer where the particle stops. The translation relations were obtained from full GEANT Monte Carlo simulations of single particles ($^3$He and $^4$He) emitted with a range of energies into the forward detector, $^3$He with 0 – 1100 MeV and $^4$He with 0 – 840 MeV. The energies were chosen, as in the experiment, such that they reached the desired forward detector planes.

*3.3 Event selection*

In order to select a clean data sample, the following criteria were used. In the first step, events with at least one charged track in the forward detector were sought. Out of these events, those corresponding to $^4$He were selected using the $\Delta E$-$\Delta E$ method (Fig. 3). Then, for each such $^4$He track, four photons or two oppositely charged pions were sought in the central detector, corresponding to the neutral and charged channels, respectively (Fig. 4).

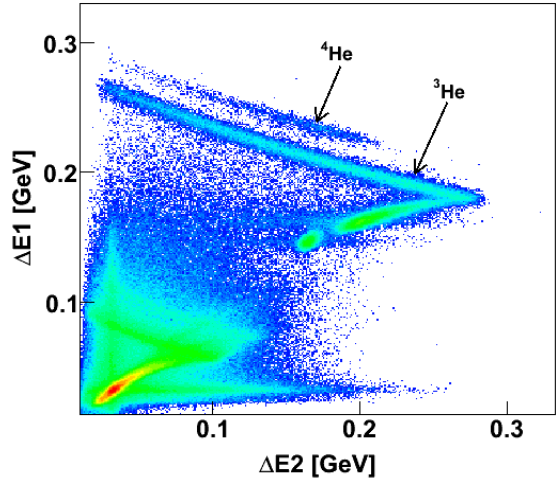

Fig. 3. $\Delta E$-$\Delta E$ plot from the forward detector at $T_d$=1029 MeV showing clear separation of $^3$He and $^4$He.

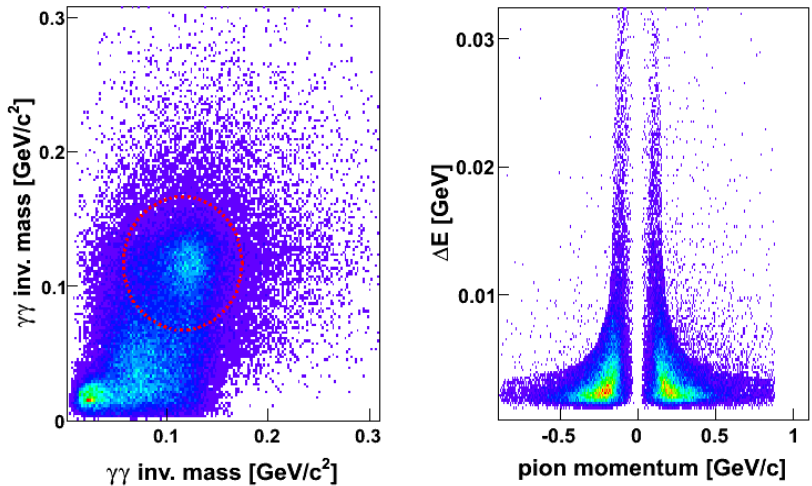

Fig. 4. Left: invariant mass of one γγ pair versus the other from the central detector calorimeter. The circle encompasses the pion events. Right: charged pion momentum in the mini-drift chamber versus the deposited energy in the plastic scintillators. The pion momentum is here multiplied by the sign of the pion charge.

In the case of the neutral channel, the best selection from the 4γ that could pair up to form two $\pi^0$ was chosen (Fig. 4 left). A kinematic fit was then applied to improve the energy resolution and to remove background. For the charged channel, the $\Delta E$-$p$ method was used, with the plastic scintillator providing the $\Delta E$ information and the mini-drift chamber the momentum $p$ (Fig. 4 right). Here also kinematic fitting was employed to improve the momentum resolution of the charged pions and to reject random background. Typically about 8-9% of the events were rejected by this procedure.

*3.4 Acceptance correction and normalization*

Since the WASA is nearly a 4π detector, the missing-mass and other distributions are fairly similar before and after acceptance correction. This also means that the final results are only weakly dependent on the model chosen for acceptance correction. We have therefore chosen isotropic phase-space distributions to calculate the acceptance correction. This should yield a good first order description of the acceptance, since the different detection efficiencies do not vary very much over the available phase space. The effects of the acceptance correction in the missing-mass and angular distributions will be demonstrated in Section 4.

The absolute normalization of the data at $T_d$ = 1029 MeV was determined by using the two-body $dd \rightarrow {}^3\text{He}\,n$ reaction, which was measured simultaneously with the $dd \rightarrow {}^4\text{He}\,\pi\pi$ reaction. A normalization factor was obtained through the comparison with the known cross sections at $T_d$ = 860 MeV and $T_d$ = 1243 MeV [33]. A plot of the differential cross section as a function of $t$-$t_{max}$, for these and our data is shown in Fig. 5, where $t$ is the four momentum transfer between the beam and the ${}^3\text{He}$ particle. The missing mass of ${}^3\text{He}$ from our data is also shown in the same figure, fitted with a Gaussian and a third degree polynomial background. The background is believed to be from nuclear interactions between the ${}^3\text{He}$ and the detector material.

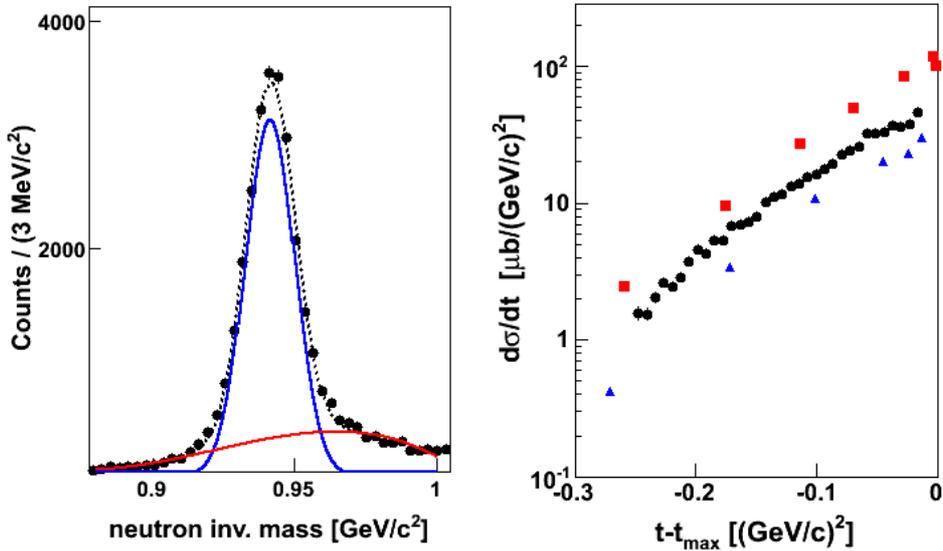

Fig. 5. Left panel: the missing mass of ${}^3\text{He}$ from our data (black dots), a 3$^{rd}$ degree polynomial background fit (red line) and a Gaussian fit (blue line) with its peak at 941.5 MeV/c$^2$ and FWHM = 20.4 MeV/c$^2$. Right panel: the $d\sigma/dt$ distribution: red squares and blue triangles (colour online, here as well as in all other figures) correspond to $T_d$ = 860 MeV and $T_d$ = 1243 MeV from Saturne measurements [33]. In addition to the small statistical errors shown by the data points, there is an absolute uncertainty of about 10%. The black circles represent our data at $T_d$ = 1029 MeV, normalized by linear interpolation between the two Saturne results.

## 4. Results and discussion

The relevant reaction channels in $dd \to {}^4\text{He}\,X$ at our beam energies of 712 MeV and 1029 MeV are:

1. $dd \to {}^4\text{He}\,\pi^o\pi^o$   2. $dd \to {}^4\text{He}\,\pi^+\pi^-$   3. $dd \to {}^4\text{He}\,\pi^+\pi^-\pi^o$

It should be noted that the $dd \to {}^4\text{He}\,\pi^o$ and $dd \to {}^4\text{He}\,\pi^o\pi^o\pi^o$ channels are forbidden by isospin conservation so that there is no background from single pion production. The $dd \to {}^4\text{He}\,\pi^o\pi^o$ and $dd \to {}^4\text{He}\,\pi^+\pi^-$ channels involve the same isospin-zero amplitude and are thus identical in all respects, except for the mass difference between the charged and neutral pions. Neglecting this mass difference, isospin invariance requires the total cross section for the charged channel to be twice as large as the neutral channel since the latter involves two identical particles. By measuring both channels with independent experimental methods for the pions, we obtain good control over unknown minor experimental effects.

Phase space distributions and the geometrical acceptance of the WASA detector for the $dd \to {}^4\text{He}\,X$ reaction at 1029 MeV and 712 MeV are shown in fig. 6.

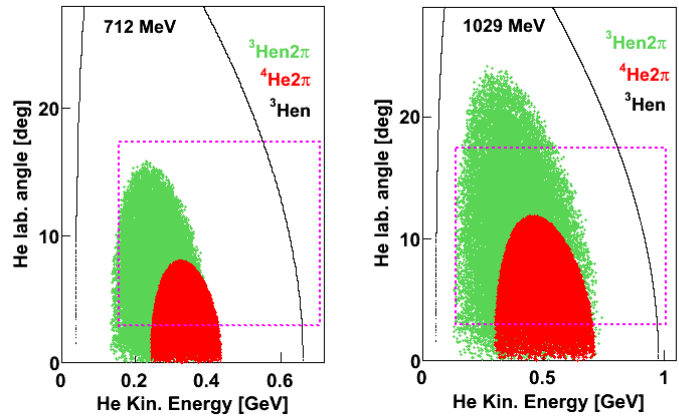

Fig.6. Phase-space simulation of the $dd \to \text{He}\,X$ reaction showing He kinetic energy versus production angle in the laboratory frame. Left: 712 MeV. Right: 1029 MeV. The dotted rectangle corresponds to the geometrical coverage of the WASA detector. The $dd \to {}^3\text{He}\,n\pi\pi$ channel (green) is the main physics background to $dd \to {}^4\text{He}\,\pi\pi$ (red) at both energies. The two-body reaction $dd \to {}^3\text{He}\,n$ (black) is used for calibration of the forward detector and for normalization of the data.

*4.1 Theoretical model*

The theoretical model with which the results are compared is a simplified version of the double-$\Delta$ GFW model [27]. In this model the $dd \to {}^4\text{He}\,\pi\pi$ reaction is treated as two independent and parallel $NN \to d\pi$ reactions taking place between projectile and target nucleons, after which the two deuterons merge together to form the ${}^4\text{He}$ (Fig. 7). In the two $NN \to d\pi$ sub-processes, the deuterons and pions are emitted preferentially at small angles with respect to the beam or target nucleon in the centre-of-mass system because of the strong p-wave dependence arising from the $\Delta$-isobar, which drives pion production. This leads to the two deuterons being emitted preferentially either parallel or anti-parallel to each other.

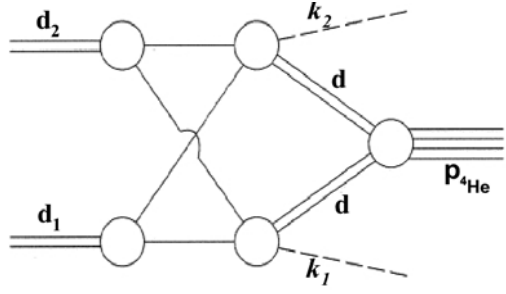

Fig. 7. Diagram for the $dd \rightarrow {}^4\text{He}\pi\pi$ reaction showing the momenta in the c.m. system [27].

When the two deuterons are emitted in the same direction, the pion pair goes in the opposite direction with small relative momentum, which translates to small ππ-invariant mass. The opposite, anti-parallel deuterons, will lead to high ππ-invariant mass. In the momentum spectra of $^4$He (cf. Fig. 8 and 9), small invariant mass corresponds to the two (forward and backward) ABC ridges and the large invariant mass to the central excess. As such, the model describes the main features of the experimental spectrum.

In the simplified version of the GFW model adopted here, only the dominant $NN \rightarrow d\pi$ amplitude, representing the S-wave coupling of a nucleon with a Δ, giving a total spin of 3/2 + 1/2 = 2, has been retained. This gives rise to the well-known $\sigma(NN \rightarrow d\pi) \propto 3\cos^2\theta_\pi + 1$ angular distribution for the pion. If only S-waves are retained for both the deuteron and $^4$He, the resulting prediction for the $dd \rightarrow {}^4\text{He}\pi\pi$ cross section is:

$$\sigma \approx FF^2 \times \left[ 3(\hat{p}\cdot\vec{k}_1)(\hat{p}\cdot\vec{k}_2) + \vec{k}_1\cdot\vec{k}_2 \right]^2 \times \text{Phase Space}, \tag{1}$$

where the $\vec{k}_i$ are pion momenta in the centre of mass and $\hat{p}$ denotes the direction of the beam. At the ABC peak, where $\vec{k}_1 \approx \vec{k}_2$, the model suggests that the angular distribution should vary as the square of the input cross section, $\sigma(dd \rightarrow {}^4\text{He}\pi\pi) \propto \left[\sigma(NN \rightarrow d\pi)\right]^2$, which is what one might expect classically.

However, the probability that the two deuterons "stick" to form the α-particle decreases away from the ABC peak. This is taken into account by the form factor, $FF$, which represents the overlap of the two internal deuterons with the $^4$He in Fig. 7. This has been evaluated numerically by integrating the $^4$He and deuteron wave functions, and the result is parameterised as [27]:

$$FF \approx \exp\left(b q_L^2 + c q_T^2 + q^2\left(d q_L^2 + e q_T^2\right)\right), \tag{2}$$

where $\vec{q} = 0.5(\vec{k}_1 - \vec{k}_2)$, $\vec{q}_L$ and $\vec{q}_T$ being its longitudinal and transverse components. The parameters have the values: $b = -0.6157$ fm$^2$, $c = -0.5415$ fm$^2$, $d = 0.0137$ fm$^4$, $e = 0.0017$ fm$^4$.

The theoretical model is used as a weight for the events generated isotropically by the same Monte Carlo code used to simulate the detector response. Thus, the theoretical model and the experimental data are compared in the same phase space covered by the detector.

*4.2 Experimental results for $dd \rightarrow {}^4\text{He}\pi\pi$ at 1029 MeV*

The raw experimental results before acceptance correction are shown in Fig. 8 and 9. Here the main features of the ABC effect are visible from a different perspective. The "rim" of the $^4$He momentum versus laboratory angle plot (top-left in each set) resembles that of a two-body reaction. This means the $^4$He and the pion pair go back-to-back in the centre of mass. Those pion pairs corresponding to $^4$He on the "rim" have small relative momentum and they correspond to the low mass enhancement in the ππ-invariant mass distribution shown at the bottom right of the figures.

Events inside the "rim", corresponding to the pion pairs having large relative momentum, make up the central bump in momentum spectrum of $^4$He or the high mass enhancement in the $\pi\pi$-invariant mass spectrum.

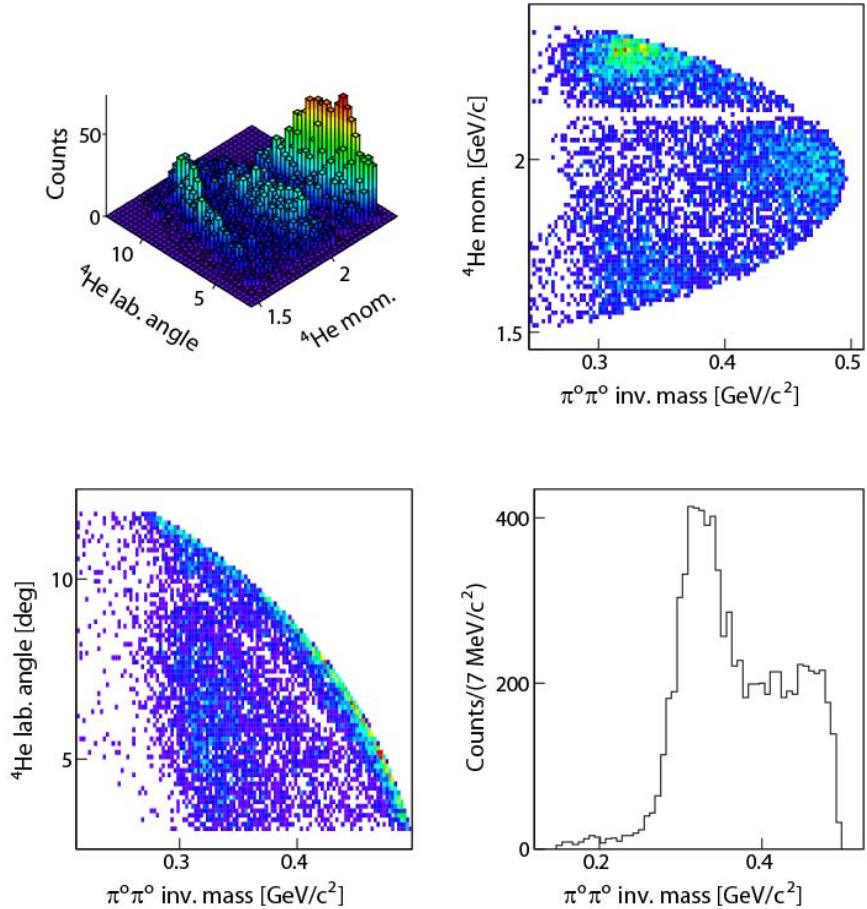

Fig. 8. Raw data from the neutral channel at 1029 MeV, before acceptance correction. Top-left: Plot of the $^4$He momentum versus production angle showing a clear outline of the ABC bands. Top-right: $^4$He momentum versus two-pion invariant mass. The depletion of events around 2.12 GeV/c is due to $^4$He being stopped in the insensitive detector region between the first and second layer of FRH. Bottom-left: The $^4$He angle versus $\pi\pi$-invariant mass. Bottom-right: The two-pion invariant mass shows the low-mass enhancement and also a slight high-mass bump.

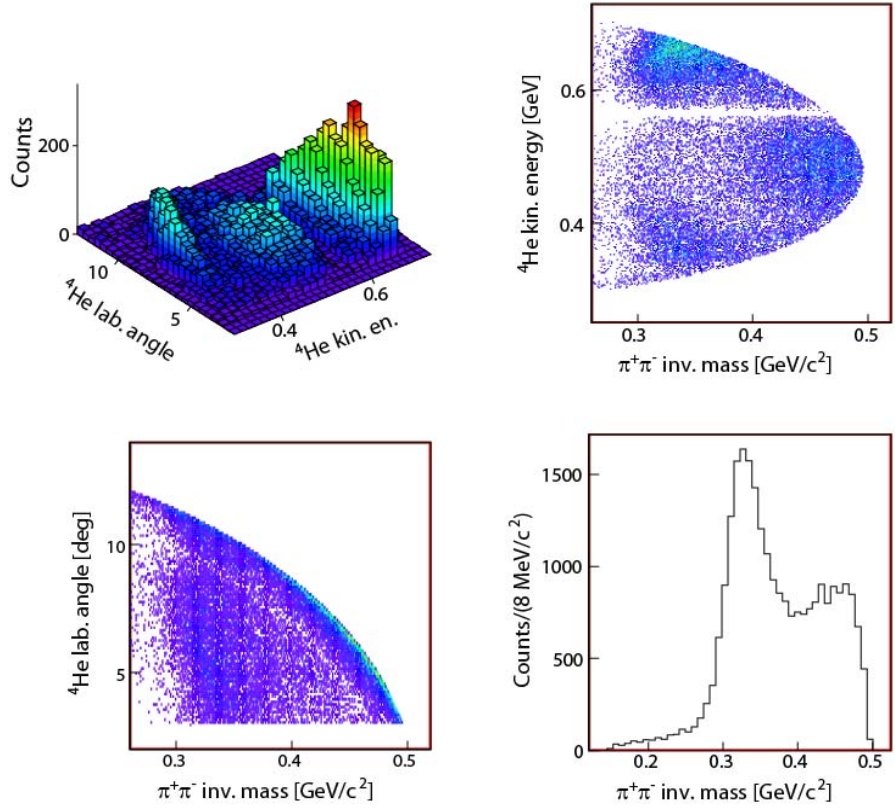

Fig. 9. Raw data from the charged channel at 1029 MeV, before acceptance correction. Top-left: two-dimensional plot of $^4$He kinetic energy versus production angle showing a clear outline of the ABC bands. Top-right: $^4$He kinetic energy versus two-pion invariant mass. Bottom-left: $^4$He angle versus $\pi\pi$-invariant mass. Bottom-right: two-pion invariant mass showing the low- and high-mass enhancements.

*4.2.1 Results after acceptance correction*

The total cross sections for the charged and neutral channels of the $dd \to {}^4\text{He}\pi\pi$ reaction at 1029 MeV were found to be $3.2 \pm 1.2$ μb and $1.9 \pm 0.8$ μb, respectively. The errors are by far dominated by the uncertainty in the normalization procedure described in Section 3.4. The errors are estimated conservatively simply by using the $dd \to {}^3\text{He}n$ data at 860 and 1243 MeV as the upper and lower limit respectively. The contributions to the error from other sources are small in comparison, e.g. errors from the 10% cross section uncertainty in the $dd \to {}^3\text{He}n$ data, the subtraction of the background under the $^3$He peak, and any difference in the acceptance for $^3$He and $^4$He.

The two-particle invariant masses (Fig. 10 and 12) and angular distributions (Fig. 13 and 14) for both neutral and charged channels are shown and compared to the theoretical predictions of the GFW model. The errors given in Figs. 10-14 as well as in later figures are statistical. Horizontal bars, where applicable, represent the bin size.

One can see from Fig. 10 that the model is in reasonable agreement with the $\pi\pi$-invariant mass data. The model reproduces nicely the relative strengths of the low mass and high mass enhancements. These experimental distributions were obtained from the independent exclusive measurements of $dd \to {}^4\text{He}\pi^o\pi^o$ and $dd \to {}^4\text{He}\pi^+\pi^-$. In Fig. 11 is shown a comparison of the missing-mass distributions from the inclusive reaction $dd \to {}^4\text{He}X$ with the direct sum of the exclusively obtained neutral and charged channel $\pi\pi$-invariant masses. This is performed as a consistency check and the results compare reasonably well, except for very small differences in the rapid rise at low invariant masses, which might be connected to the kinematic fitting imposed on the exclusive measurements.

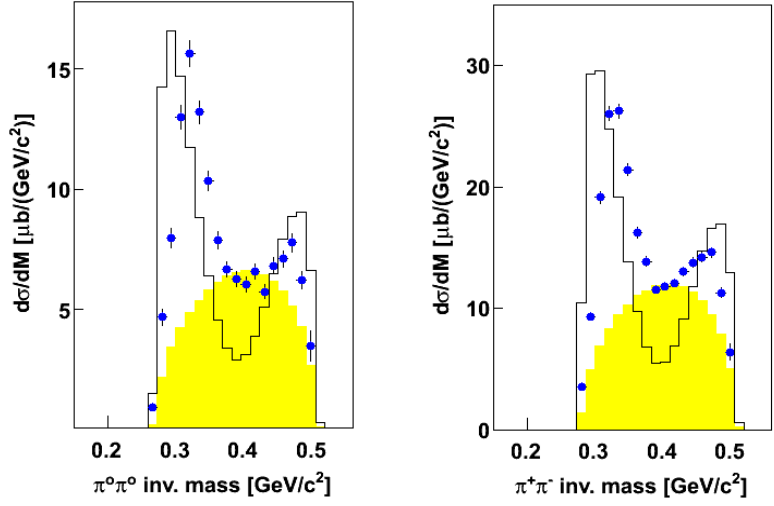

Fig. 10. ππ-invariant mass for the neutral (left) and charged (right) channels at $T_d$=1029MeV. Filled circles are from experiment, the full drawn histogram is the GFW model [27] and the shaded histogram is isotropic phase space, arbitrarily normalized. The GFW model, as adopted here, is in qualitative agreement with the experimental results.

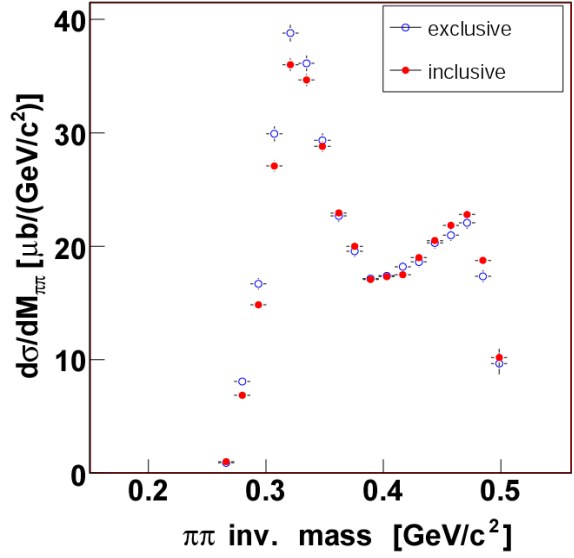

Fig. 11 Comparison of the ππ invariant-mass distributions obtained from the sum of the direct neutral and charged channel measurements (exclusive sum, unfilled circles) and the missing mass from the inclusive $dd \rightarrow {}^4\text{He}\,X$ measurement (inclusive, filled circles).

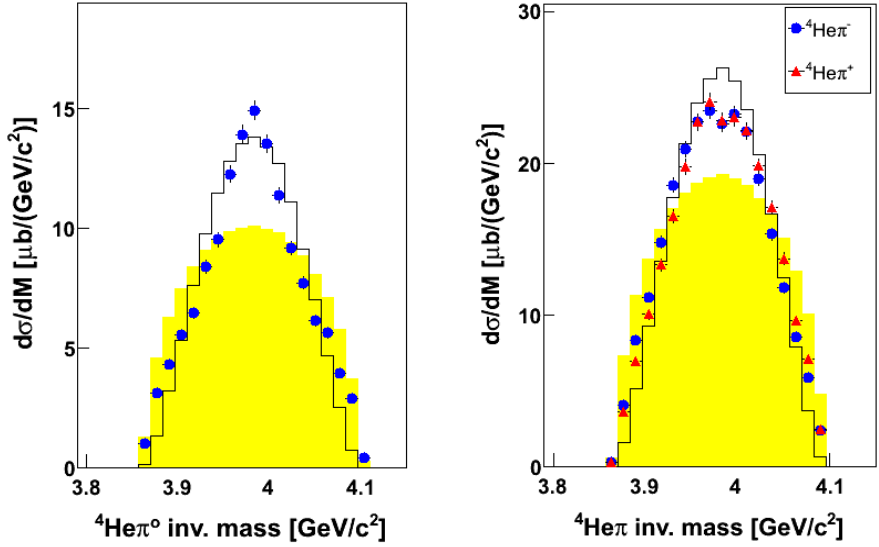

Fig. 12. $^4$He$\pi$ invariant masses for the neutral (left) and charged (right) channels at T$_d$ =1029 MeV. Filled points are from experiment, the full drawn histogram corresponds to the GFW model, and the shaded histogram is isotropic phase space. Here the GFW model and phase-space are normalized with respect to data so that the model and phase space have the same area as data. In the right panel, the circles correspond to $^4$He$\pi^-$ and the triangles to $^4$He$\pi^+$.

In the $^4$He$\pi$ invariant mass plots of Fig. 12, the agreement between data and model is very good for both channels.

In the angular distributions of the $^4$He and pions (Fig. 13 and 14), the agreement between data and model is reasonable. The pions show strong forward-backward peaked angular distributions, while the $^4$He distribution is rather flat in the middle. Note that all angular distributions in $dd \to {}^4\text{He}\,\pi\pi$ have to be forward-backward symmetric in the centre-of-mass system and this is fairly well borne out by the experimental results. Any slight residual asymmetry in the experimental results is probably due to small imperfections in the Monte Carlo description of the detector.

In both the neutral and charged channels, the angular distribution of a pion in the $\pi\pi$-subsystem (Jackson frame) can be extracted. The angle here is that between the pion momentum and that of the incoming deuteron beam in the $\pi\pi$ rest frame. The angular distributions are presented in the third column of Figs. 13 and 14, and they show a significant dependence on the $\pi\pi$-invariant mass. The $\cos\theta_\pi^{\pi\pi}$ distribution is forward-backward peaked when all $\pi\pi$-invariant masses are included, but has a central maximum for events with $\pi\pi$-invariant masses lower than 360 MeV/c$^2$. Such behaviour is predicted by the GFW model. This dependence can also be inferred from the two-dimensional plot of Fig. 15, where the $\pi\pi$-invariant mass is shown versus $\cos\theta_\pi^{\pi\pi}$ for charged pions.

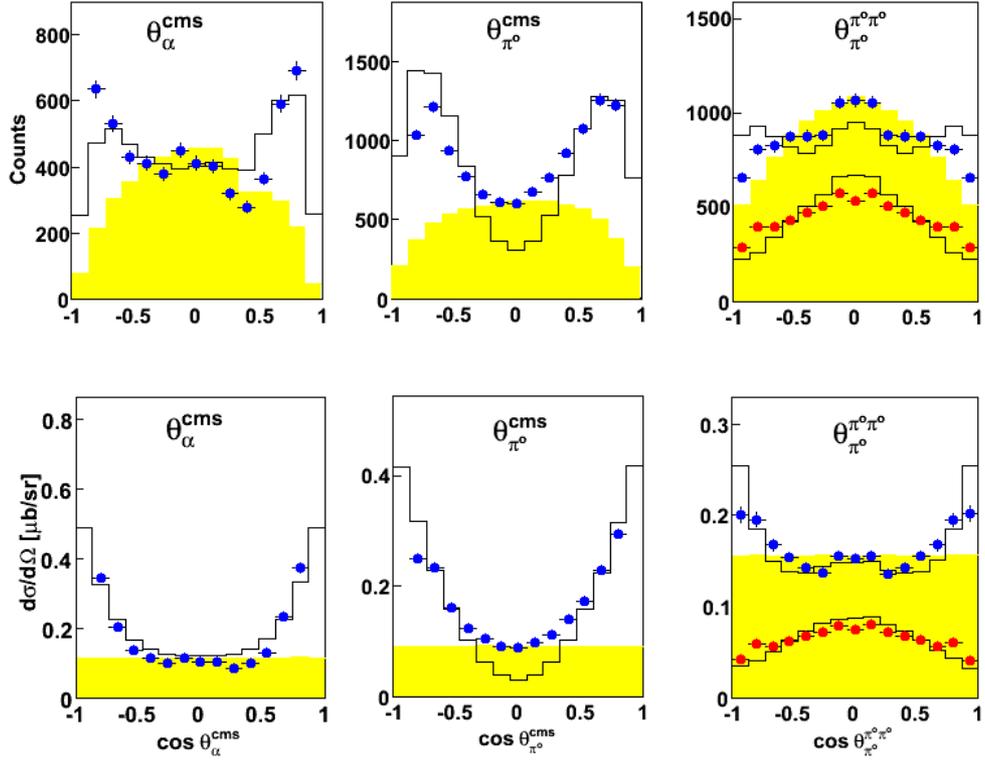

Fig. 13. Angular distribution of $^4$He and neutral pions at $T_d$=1029 MeV before (top-row) and after (bottom-row) acceptance correction. Filled points are experimental, the black histogram corresponds to the GFW model, and the shaded histogram is isotropic phase space, arbitrarily normalized. The red points (right hand column, lower group of points) are from the region of $\pi^o\pi^o$-invariant masses below 360 MeV/c². The noteworthy features here are the strong angular dependence of $^4$He and the pions (first and middle column) as well as the non-isotropic angular distribution of the $\pi^o$ in the $\pi^o\pi^o$-subsystem (third column). The GFW model describes well the difference in shape of this distribution for the two $\pi^o\pi^o$-invariant mass region.

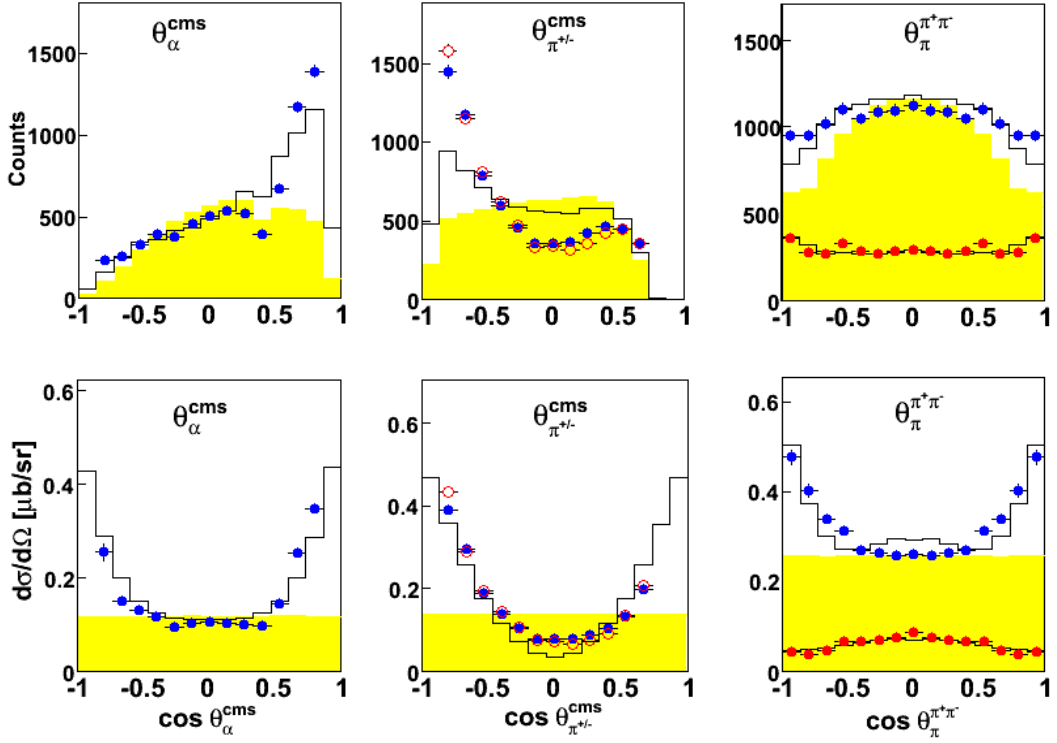

Fig. 14. Same as Fig. 13 but for the charged pion channel. In the middle graph the $\pi^+$ and $\pi^-$ distributions are shown separately. The non-filled red points denote $\pi^+$ and the filled blue points denote $\pi^-$.

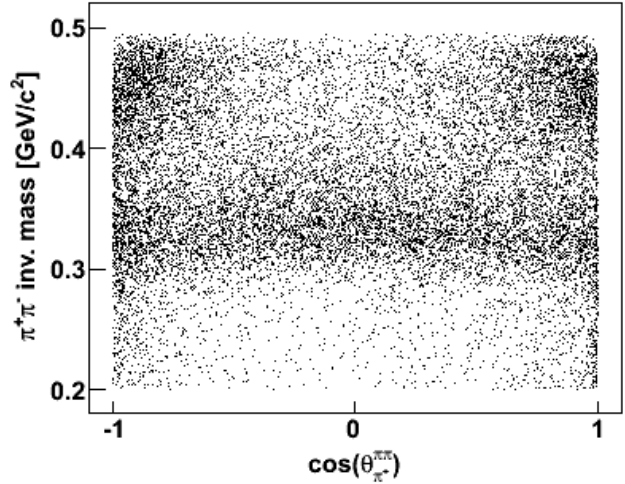

Fig. 15. Angular dependence of a pion in the ππ-subsystem (Jackson frame) versus the ππ-invariant mass for the charged channel. It can be clearly seen that the forward-backward steepness in the angular distribution of pions in the ππ-subsystem comes dominantly from the high mass enhancement region, where, in the overall cm system the pions are emitted almost back-to-back and at small angles with respect to the beam axis. The pions in the low mass enhancement are almost "isotropic" in this sense. This remarkable behaviour, which is well reproduced by the GFW model, is similar for the neutral channel.

It should finally be emphasized that the spectra obtained for the invariant mass and angular distributions in the $dd \to {}^4\text{He}\pi\pi$ reaction have common features with those obtained for the $pd \to {}^3\text{He}\pi\pi$ case [5], and even to a large extent also for the $pd \to p_{sp}d\pi^0\pi^0$ case [11].

*4.3 The $dd \to {}^4\text{He}\pi^+\pi^-\pi^o$ background contribution to the $dd \to {}^4\text{He}\pi^+\pi^-$ events*

In the analysis of the $dd \to {}^4\text{He}\pi^+\pi^-$ reaction it has been assumed that the cross section for three-pion production is very small. This can be verified from a study of the missing-mass distributions shown in Fig. 16.

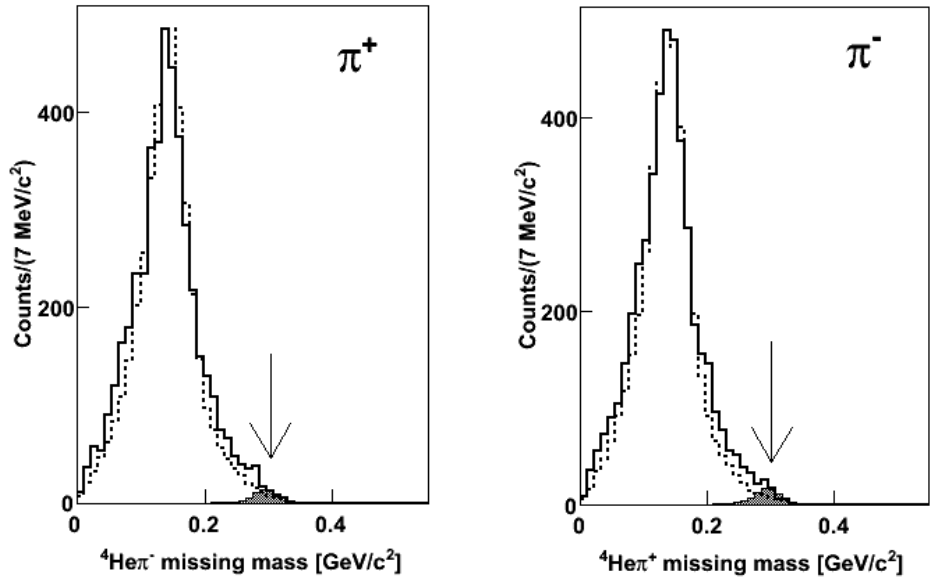

Fig. 16. The missing mass of ${}^4He\,\pi^-$ (left) and ${}^4He\,\pi^+$ (right). The bold lines correspond to data. The dashed line is $dd \to {}^4\text{He}\pi^+\pi^-$ from Monte Carlo calculations and scaled to the peak of the experimental result, the momenta of ${}^4$He and pions being smeared by 1% and 7%, respectively. The shaded area is the resulting histogram from a simulation of $dd \to {}^4\text{He}\pi^+\pi^-\pi^o$, scaled to coincide with the experimental histogram. The shaded area represents less than 2% of the data.

In previous inclusive experiments, it might have been argued that the central bump in the ${}^4$He momentum spectra included significant contributions from the $3\pi$ and $\eta$ production channels [14]. In $dd \to {}^4He\,X$ at $T_d = 1029$ MeV, however, the $dd \to {}^4\text{He}\pi^+\pi^-\pi^o$ cross section shown in Fig.16 is too small to account for the central bump. It should be noted that $dd \to {}^4\text{He}\pi^o\pi^o\pi^o$ is not allowed by isospin conservation.

*4.4 Experimental results for $dd \to {}^4\text{He}\pi\pi$ at 712 MeV*

The experimental results discussed thus far correspond to the beam energy of 1029 MeV. The experimental conditions at 712 MeV were much poorer and only semi-quantitative conclusions may be drawn. Experimental plots of the ${}^4$He production angle versus kinetic energy at $T_d = 712$ MeV (left panel) and 1029 MeV (right panel) are compared in Fig. 17. Despite the background, one can still see in the left panel an outline that is similar in features to that on the right. These are signatures of the ABC effect, corresponding to the low-mass and high-mass enhancements in the $\pi\pi$-invariant mass distributions, as discussed in section 4.2 (Fig. 8 and 9).

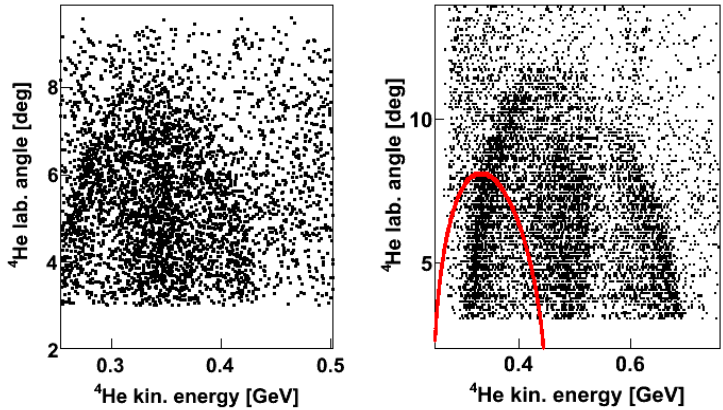

Fig. 17. Laboratory production angle versus kinetic energy of $^4$He at 712 MeV (left) and at 1029 MeV (right), both for the neutral channel; the red line shows the limits of the region accessible at 712 MeV.

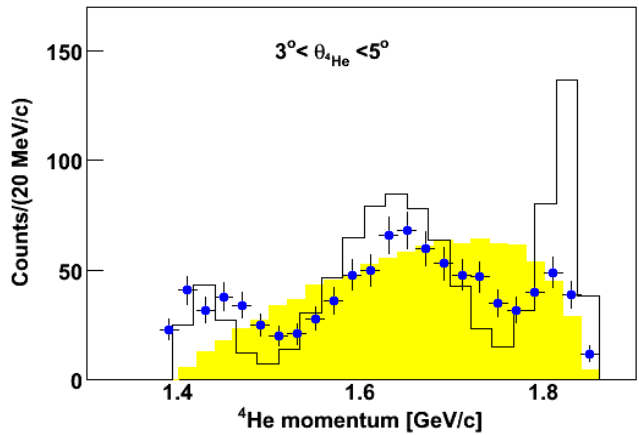

Fig. 18. The $^4$He momentum is shown for a slice in $^4$He laboratory angle at 712 MeV, before acceptance correction. The dots are from experiment while the histogram and shaded area correspond to the GFW model and pure phase space, respectively. The typical three-peak signature of the ABC effect is clearly visible though, from the forward-backward symmetry, it seems that events are being lost at high momentum.

This result, other than confirming the ABC effect at the lowest energy yet, also shows that the ABC phenomenon in $dd \to \,^4\text{He}\,\pi\pi$ consists of both a low mass and a high mass enhancements, as can be seen from the two peaks and the large central bump in the momentum spectra of $^4$He in Fig. 18. On the other hand, since the cross section has to be symmetric in the centre of mass, it appears that events are being lost at high $^4$He momentum. Note that the energy of 712 MeV is well below the threshold for $dd \to \,^4\text{He}\,\pi\pi\pi$ so that only two-pion production is allowed.

## 5. Conclusions

The ABC effect has been studied exclusively in the $dd \to {}^4\text{He}\,\pi^+\pi^-$ and $dd \to {}^4\text{He}\,\pi^0\pi^0$ reactions using the CELSIUS storage ring with the WASA detector assembly. Due to their isospin selectivity, these channels are expected to show the most prominent enhancements. Inclusive measurements had already demonstrated dramatic structures for all beam energies between 800 MeV and 2.4 GeV, with a central bump as well as forward and backward ABC peaks. Our experiment was carried out at two energies, viz $T_d = 712$ MeV, which is so far the lowest energy at which the ABC effect has been observed, and at $T_d = 1029$ MeV, where the ABC effect is at its strongest but where exclusive measurements were not available.

At 1029 MeV the total cross sections for the neutral and charged channels were determined to be $1.9 \pm 0.8$ µb and $3.2 \pm 1.2$ µb, respectively, the major uncertainty in both cases arising from that in the overall normalization factor. At both 1029 and 712 MeV one sees the characteristic ABC structure of a sharp enhancement close to the minimum two-pion mass and a rather broader one near the maximum invariant mass. In earlier inclusive experiments, contributions from three pions or even η-production could not be conclusively ruled out. However, in the exclusive measurement at 1029 MeV, it is shown that the $dd \to {}^4\text{He}\,\pi^+\pi^-\pi^o$ contribution alone is much too small to account for any significant part of the high mass enhancement in the charged channel. Furthermore, the same high mass enhancement is observed in the neutral channel, where the $dd \to {}^4\text{He}\,\pi^o\pi^o\pi^o$ channel is forbidden by isospin invariance. The beam energy is also below the η-production threshold.

Both the low and high mass enhancements in the ππ spectrum are reasonably well reproduced by the simplest version of the model proposed by Gårdestig, Fäldt, and Wilkin [27]. This model, which explains the ABC effect in this reaction as a kinematic enhancement arising from the independent production of two *p*-wave pions, is in principle parameter-free, though we have here normalized the predictions to the integrated cross sections.

Angular distributions of the charged and neutral pions are presented for the first time. The strong angular dependence of both the pions and the α-particle in the centre of mass frame follow closely the GFW predictions including a change of the pion angular distribution in the ππ rest frame that goes from concave to convex, depending upon the value of ππ-invariant mass. It is therefore evident that the ABC enhancement in $dd \to {}^4\text{He}\,X$ is qualitatively consistent with a kinematic effect that owes its origins to the double p-wave term contained in the matrix element of Eq. (2). However, other possible explanations of the ABC enhancement are not excluded. To go further with the GFW approach, and also to establish the normalization, it would be necessary to implement the full GFW model, which would require the introduction also of the small $NN \to d\pi$ amplitudes into the calculation [27].

Exclusive experiments would also be welcome at lower energies in order to further test the model and see how the ABC effect fades away when approaching threshold [15,16].

**Acknowledgements**

This was the final experiment carried out at the CELSIUS machine and we are grateful for the support of the personnel at The Svedberg Laboratory and in particular D. Reistad. Special thanks are directed to C.-J. Fridén and G. Norman for their unfailing and skilful work on the pellet target. Discussions with G. Fäldt on the double-Δ model have been very helpful. This work was supported by the European Community under the "Structuring the European Research Area" Specific Programme Research Infrastructures Action (Hadron Physics, contact number RII3-cT-204-506078), by the Swedish Research Council and by BMBF (06TU201).